\documentclass[twocolumn,aps,prl,showpacs,superscriptaddress]{revtex4}

\usepackage{subfigure}
\usepackage{graphicx}
\usepackage[intlimits]{amsmath}
\usepackage{amsfonts}
\usepackage{amssymb}
\usepackage{amsmath}
\usepackage{amscd}
\usepackage{latexsym}
\usepackage{bbm}
\usepackage[sort&compress]{natbib}
\usepackage{ifthen}
\usepackage{epsfig}
\usepackage{dcolumn}      
\usepackage{bm}           
\usepackage{amsfonts}
\usepackage{enumerate}
\usepackage[latin1]{inputenc} 

\newcommand{\bibi}[6]{\bibitem{#1}#2, \emph{#3} \textbf{#4}, #5 (#6)}

\begin{document}
\title{Enhanced Transmission and Reflection of Femtosecond Pulses by a Single Slit}



\begin{abstract}
We show that a physical mechanism responsible for the enhanced transmission and reflection of femtosecond pulses by a single subwavelength nanoslit in a thick metallic film is the Fabry-Perot-like resonant excitation of stationary, quasistationary and nonstationary waves inside the slit, which leads to the field enhancement inside and around the slit. The mechanism is universal for any pulse-scatter system, which supports the stationary resonances. We point out that there is a pulse duration limit below which the slit does not support the intraslit resonance.
\end{abstract}


\author{M. Mechler}
\affiliation{South-Trans-Danubian Cooperative Research Centre, University of P\'ecs, Ifj\'us\'ag u.\ 6., 7624 P\'ecs, Hungary}

\author{O. Samek}
\affiliation{Institute for Analytical Sciences, Bunsen-Kirchhoff-Strasse 11, 44139 Dortmund, Germany}

\author{S.V. Kukhlevsky}
\affiliation{Department of Physics, University of P\'ecs, Ifj\'us\'ag u.\ 6., 7624 P\'ecs, Hungary}

\maketitle

In the last decade a great number of studies have been performed to map the effects and potential applications of enhanced scattering of continuous light waves by metallic subwavelength nanoapertures (for example, see the studies \cite{ebbesen,barn} and references therein). With the fast development of ultrashort-pulse lasers naturally arose the question: How does a subwavelength aperture affect a femtosecond (fs) light pulse? Experiments as well as computations  were performed to investigate this problem for various parameters: pulse length, object shapes and sizes, etc.\cite{nechay,mitr,stoc,pack,stav,kukh1,dechant,muller,labardi,bori,ben,neerhoff,betzig} It was shown that the scattering process alters the pulse shape and spectrum, resulting in various effects such as pulse broadening, compression, time delay and advancement, and spectral shifts. The recent study\cite{kukh1} showed that the transmission of fs pulses through a single slit  can be enhanced at appropriate resonant conditions. Although a general point of view has been suggested on the physics underlying the enhanced transmission and reflection of fs pulses by various nanostructured materials, as due to excitation of the long-lived quasistationary eigenstates of the effective Hamiltonian for the Maxwell theory~\cite{bori}, a detailed investigation of the problem for the most fundamental system of wave physics (a single slit) is required.

In the present paper, a direct solution of Maxwell's equations shows that a physical mechanism responsible for the enhanced scattering (transmission and reflection) of fs pulses by a single subwavelength nanoslit in a thick metallic film is the Fabry-Perot-like resonant excitation of Fourier components (stationary waves) of a fs wavepacket inside the slit. That leads to the resonant excitation and enhancement of stationary, quasistationary and nonstationary fs pulses inside the slit with the enhanced field around the slit. The mechanism is universal for any pulse-scatter system that supports the stationary resonances. We point out that there is a pulse duration limit below which the slit does not exhibit the enhanced scattering. In such a case the intraslit fields are quite different from those observed in the case of stationary (continuous) waves.

\begin{figure}[hb]
\includegraphics[keepaspectratio,width=9cm]{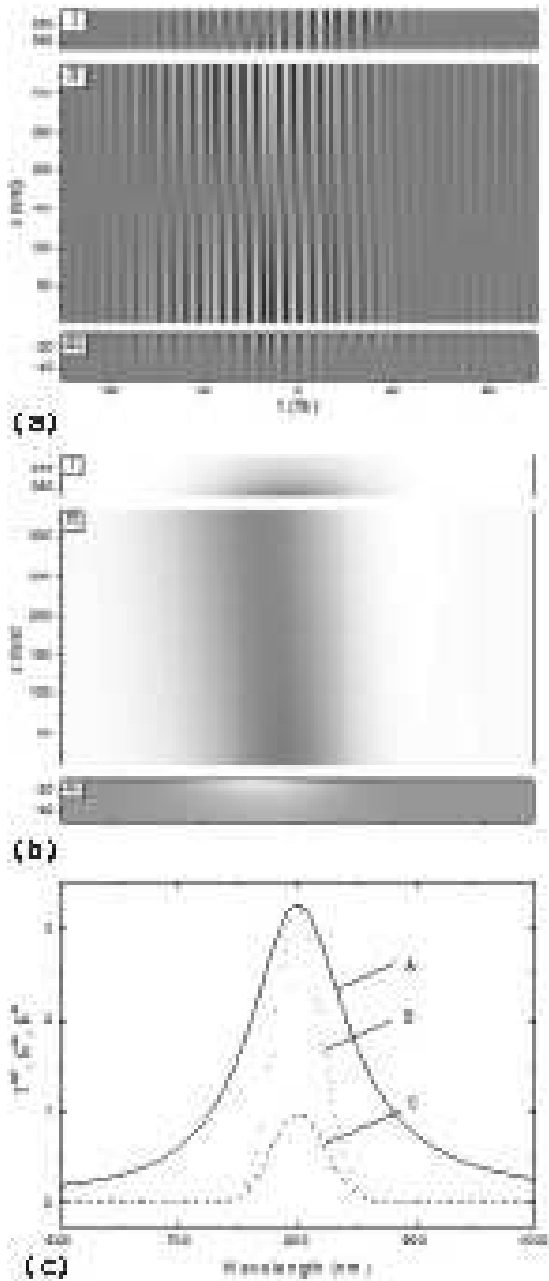}
\caption{\label{fig:25fs} The stationary regime of enhancement of
a 25~fs Gaussian pulse in front(I), inside (II) and behind (III) a
subwavelength slit. (a) Electric field distribution
$\mathrm{Re}(E_x)$. (b) Energy flux $S_z$. (c)
Amplitude transmission $T^{1/2}$ (A), spectra of transmitted 
$E^{tr}$ (B) and incident $E^{in}$ (C) pulses.}
\end{figure}

\begin{figure}[hb]
\includegraphics[keepaspectratio,width=9cm]{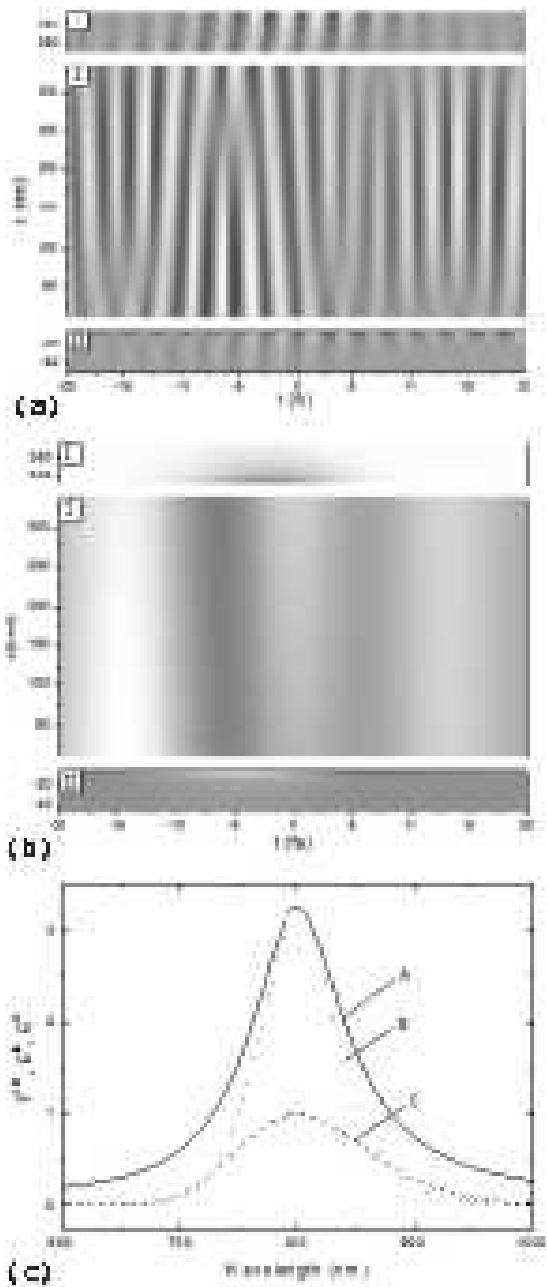}
\caption{\label{fig:10fs} The quasistationary regime of
enhancement of a 10~fs Gaussian pulse in front (I), inside (II)
and behind (III) a subwavelength slit. (a) Electric field
distribution $\mathrm{Re}(E_x)$. (b) Energy flux $S_z$. (c)
Amplitude transmission $T^{1/2}$ (A), spectra of transmitted 
$E^{tr}$ (B) and incident $E^{in}$ (C) pulses.}
\end{figure}

\begin{figure}[hb]
\includegraphics[keepaspectratio,width=9cm]{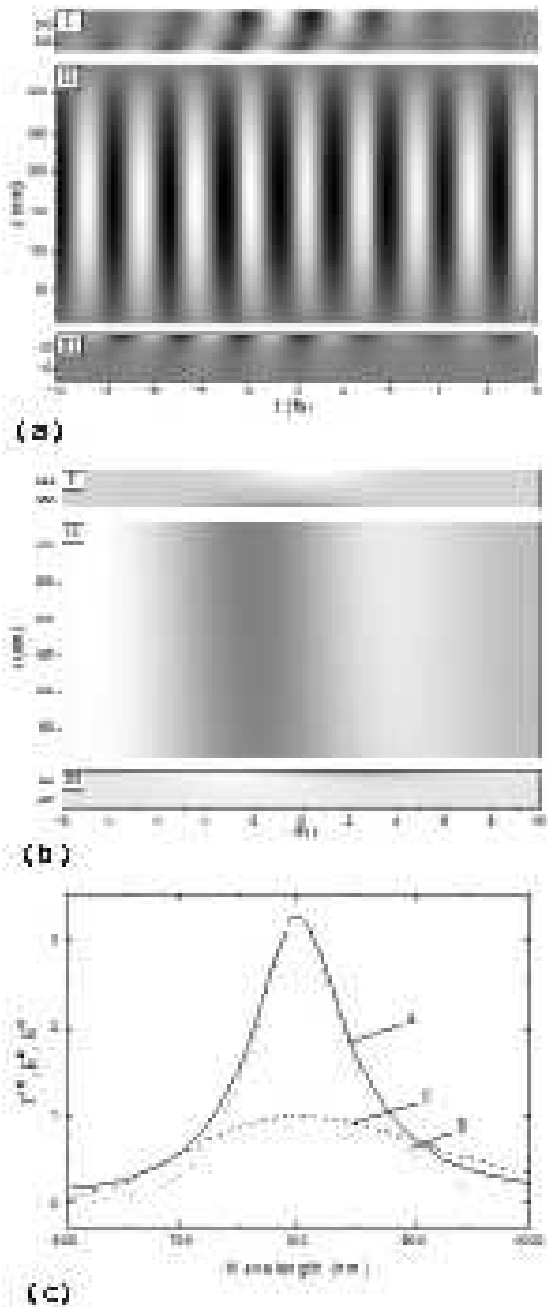}
\caption{\label{fig:5fs} The nonstationary regime of enhancement
of a 5~fs Gaussian pulse in front (I), inside (II) and behind
(III) a subwavelength slit. (a) Electric field distribution
$\mathrm{Re}(E_x)$. (b) Energy flux $S_z$. (c)
Amplitude transmission $T^{1/2}$ (A), spectra of transmitted 
$E^{tr}$ (B) and incident $E^{in}$ (C) pulses.}
\end{figure}

The model used in the present study is based on the Neerhoff and Mur rigorous numerical solution of Maxwell's equations for a continuous wave (CW) passing through a subwavelength slit in a thick metallic screen of perfect conductivity \cite{neerhoff}. As a light pulse can be treated mathematically like the sum of Fourier components (CW's) of the wavepacket, we extended this model to Gaussian fs pulses through the Fourier decomposition. The incident pulse has been first decomposed into a sum of CW's through a discrete Fourier transform. The Neerhoff and Mur method then was applied to these continuous waves (for details, see Ref. \cite{betzig}). Finally, the inverse Fourier transform was applied to the CW's to restore the pulse. The analysis was performed for different parameters of the pulse-slit system. The three general regimes of the pulse enhancement have been found, which correspond to long, medium and short incident pulses. The results presented in Figs. 1-3 demonstrate the regimes for the following parameters of a pulse-slit system: central wavelength of the wavepacket $\lambda=800$~nm (the wavelength of a Ti:Sapphire laser), slit width $2a=25$~nm, screen thickness $b=345.4$~nm. The values of $\lambda$, $a$ and $b$ correspond to the resonant conditions for enhancement of CW according to Ref. \cite{kukh1}. The pulse (${\vec{H}}(x,y,z,t)=U(x,z){\exp[-2\ln(2)(t/{\tau})^2]}{\exp}(-i\omega_{0}{t}){\vec{e}}_y$, where $\tau$ is the pulse duration and $\omega_0=2{\pi}c/\lambda_{0}$ is the central frequency) was assumed to be TM polarized; the pulse was considered to fall normally onto the slit. The computation results are shown for three different pulse durations: 25~fs, 10~fs and 5~fs (Figs. \ref{fig:25fs}, \ref{fig:10fs}, and  \ref{fig:5fs}, respectively). The electric field distribution $E_x$ (Figs. 1(a), 2(a) and 3(a)) and the flux $S_z$ (Figs. 1(b), 2(b), and 3(b)) are shown as the function of time $t$ and position on the $z$ axis. The time evolutes from positive values towards the negative values (in the figures, from the right to left), while the pulse propagates in the $z$ direction from the positive values towards the negative values, i.e. from the top towards bottom.

In the case of a long (25~fs) pulse, which corresponds to a narrow
spectral width of the wavepacket, practically all the Fourier
components experience the intra-slit resonant enhancement. At any
time during the pulse, the real part of the electric field spatial
distribution (Fig.~\ref{fig:25fs}(a)) appear to be quite similar
to the resonances observed with a CW\cite{kukh1}, except that it
has a temporal distribution. The enhancement regime shown in
Fig.~\ref{fig:25fs}(a) can be called as stationary one. In such a
regime, the amplitude of the transmitted pulse is given with high
accuracy by $E^{tr}=E^{in}\sqrt{T_{CW}}$, where
$T_{CW}=S_z^{tr}/S_z^{in}$ is the transmission coefficient. Only a
very small difference can be observed. It can be explained by the
fact that the phase shifts of the different frequency components
at the slit entrance and exit are approximately the same. Inside
the slit, the maximum of the distribution near the input and
output of the slit are slightly shifted from that of
CW~\cite{kukh1}. The effect is caused by the internal reflection
of the pulse at the slit entrance and exit. We also observe the
interference of the incident, reflected and diffracted fields at
the slit entrance. A pronounced zero level is produced between the
incident and reflected fields that move with the time. Notice that
in front of the slit, the diffracted field has a significant
effect on the interference only in the near-field region. Behind
the slit, one can see the output pre-pulse (at $\approx$20 fs) and
post-pulse between -20 and -40 fs. In Fig.~\ref{fig:25fs}(b), the
pulse shape shows two important features: a pulse delay of
$\approx 8$~fs (theoretical transmission time is $\approx
$~1.15~fs) and also the output pre-pulse and post-pulse. The pulse
delay is in agreement with that observed in the case of thick
screens~\cite{stav}. However, it is different from the
superluminarity predicted for thin screens~\cite{mitr,kukh1}.
Notice that the output prepulse and postpulse have an energy flow
direction opposite to the main pulse. The maximal value of the
energy flux behind the slit is approximately half of that inside
the slit and 10/1 of that in front of the slit. The transmission
coefficient is given with high accuracy by $T=T_{CW}=10$. Analysis
of Figs.~\ref{fig:25fs}(a) and (b) shows the pulse enhancement and 
temporal broadening and delay, while Fig.~\ref{fig:25fs}(c) indicates 
the amplification of the spectral components, and the spectral narrowing
(temporal broadening: $\tau^{tr}/\tau^{in}=1.15$) of the transmitted pulse.

In the case of a shorter (10 fs) pulse having a more wide spectra of the wavepacket, the enhancement regime is different (Fig.~\ref{fig:10fs}(a)) from the stationary regime observed in the case of a long (25~fs) pulse. In this regime, most of the Fourier components experience the intra-slit resonant enhancement. Concerning the phase shift, most of the frequency components endure nearly the same shift at the slit entrance and exit as the central frequency. However, in this case there are some components that have completely different shift from that of the central frequency. Although the field distribution is periodically reproduced during the pulse, the distributions are altered from that of stationary distributions. The amplitude of transmitted pulse is smaller than in the previous long-pulse case ($E^{tr}<E^{in}\sqrt{T_{CW}}$). The enhancement regime shown in Fig.~\ref{fig:10fs}(a) can be called as quasistationary one. We notice that the intra-slit resonances show a more pronounced multiple reflection. The reflections occur in approximately 6 fs, they can be attributed to the long living resonances observed also in the study~\cite{bori}. This effect is attributed to the interference of the waves (Fourier components) having slightly different phases inside the slit. Fig.~\ref{fig:10fs}(b) shows the energy flux of the 10~fs pulse.  We notice a pulse delay of $\approx 5$~fs and the internal reflection (a dark area at $t=-5$~fs). The fact of the internal reflection is supported by the time difference ($b/c\approx 1.15$~fs) between the minimum values at the two ends of the slit. The maximum value of the energy flux behind the slit is nearly equal to that inside the slit and $\approx 4/1$ of that in front of the slit. The flux enhancement is somewhat smaller than in the stationary case, $S_z^{tr}=T_{CW}S_z^{in}$. Figures~\ref{fig:10fs}(a) and (b) show the pulse enhancement and temporal broadening and delay, while Fig.~\ref{fig:10fs}(c) indicates
the amplification of the spectral components, and the spectral narrowing (temporal broadening: $\tau^{tr}/\tau^{in}=1.64$) of the transmitted pulse.

Comparison of the electric field distributions of a 5~fs pulse with a 25~fs and 10~fs pulses (Fig.~\ref{fig:5fs}(a),  Fig.~\ref{fig:25fs}(a) and Fig.~\ref{fig:10fs}(a), respectively) reveals the most pronounced difference. In this enhancement regime, considerable part of the Fourier components does not experience the intra-slit resonant enhancement.
Examining again the phase shift, the majority of the frequency components suffer quite different shifts from that of the central frequency. Although the field distribution is periodically reproduced during the pulse, the distributions are completely different from that of stationary and quasi-stationary distributions. The very weak enhancement regime shown in Fig.~\ref{fig:5fs}(a) can be called as nonstationary one. Indeed, in Fig.~\ref{fig:25fs}(a) the instantaneous distribution is similar to that observed with a CW \cite{kukh1}, i.e., minimum at one end of the slit and maximum at the other, while Fig.~\ref{fig:5fs}(a) shows one maximum approximately at the center of the slit. We also notice a wave-like fluctuation in the time of the maximum relative to the middle of the slit. This effect is caused by the interference of the waves (Fourier components) having different phases inside the slit. Fig.~\ref{fig:5fs}(b) shows the energy flux of the 5~fs pulse. We notice that the flux is completely different from that of a 25-fs pulse. In this case, the maximal value of the energy flux behind the slit is approximately equal to the value observed inside the slit, and $\approx 1.3$ times bigger than that in front of the slit. Figures~\ref{fig:5fs}(a) and (b) show the pulse enhancement and temporal broadening and delay, while Fig.~\ref{fig:5fs}(c) indicates 
the amplification of the Fourier components, and the spectral narrowing (temporal broadening: $\tau^{tr}/\tau^{in}=1.15$) of the transmitted pulse.

In conclusion, a direct solution of Maxwell's equations showed that a physical mechanism responsible for the enhanced scattering (transmission and reflection) of fs pulses by a single subwavelength nanoslit in a thick metallic film is the Fabry-Perot-like resonant excitation of Fourier components (stationary waves) of a fs wavepacket inside the slit. That leads to the resonant excitation and enhancement of stationary, quasistationary and nonstationary fs pulses inside the slit with the enhanced field around the slit. The three general regimes of the pulse enhancement have been found, which correspond to long, medium and short incident pulses. The enhancement mechanism is universal for any pulse-scatter system that supports the stationary resonances. We point out that there is a pulse duration limit below which the slit does not exhibit the enhanced scattering. In such a case the intraslit fields are quite different from those observed in the case of stationary (continuous) waves. We believe that the presented results clarify not only the physical mechanism of the enhanced scattering of fs pulses by a slit (the most simple and fundamental system of wave physics), but gain physical inside into the nature of resonant enhancement of fs pulses by more complex nanosystems.

\end{document}